# Millimeter-level Resolution Photonic Multiband Radar Using a Single MZM and Sub-GHz-Bandwidth Electronics

**Peixuan Li\*, Wenlin Bai, Xihua Zou, Ningyuan Zhong, Wei Pan, and Lianshan Yan**
*Center for Information Photonics and Communications, School of Information Science and Technology, Southwest Jiaotong University, Chengdu 611756, China*
*\*Corresponding author: lipeixuan@swjtu.edu.cn*

**Abstract:** We here propose a novel cost-effective millimeter-level resolution photonic multiband radar system using a single MZM driven by a 1-GHz-bandwidth LFM signal. It experimentally shows an ~8.5-mm range resolution through coherence-processing-free multiband data fusion.

## 1. Introduction

Radar systems of ultrahigh-resolution target detection and imaging are of great importance in civilian and military applications. However, the inherent dependence of high detection resolution on the large instantaneous bandwidth imposes great challenges on state-of-the-art electronics. Fueled by the unique advantages of broad bandwidth, low frequency-dependent loss, immunity to electromagnetic interference, and fast analog signal processing, microwave photonics (MWP) technologies have gained more and more momentum to overcome the bandwidth and operation frequency limitations of traditional electronic radar systems [1]. Using broadband photonic signal generation schemes like optically-injected semiconductor laser [2] and optical frequency-shifting loop [3], millimeter-level resolution radar detection is achieved. However, their bulky system structure and complex bias control greatly hinder their real-world deployments. On the other hand, multiband radar fusion offers a viable solution to achieve an ultrahigh range resolution using low-speed electronics as well as extraordinary versatility to accommodate various operation conditions [4-6]. In particular, MWP technologies are distinguished by their huge bandwidth and flexibility to generate and process multiband radar signals [5] in a hardware-efficient manner. In [6], millimeter-level range resolution detection and imaging are achieved thanks to the fusion of three 2-GHz-bandwidth MWP radars. Whilst, the requirement of multiple parallel bulky photonic transceivers working in disjoint frequency bands raises significant concerns for the cost, size, weight, and power (SWaP) of the system. Moreover, the absence of coherence between radar subbands demands heavy computational efforts for cohering measurement data of different frequency bands [4,6].

In this work, we propose and experimentally demonstrate a novel compact and cost-effective millimeter-level resolution photonic multiband radar system. It is based on a single Mach-Zehnder modulator (MZM) driven by a large-amplitude intermediate-frequency (IF) linear frequency-modulated (LFM) signal. Harmonic signals from the modulation nonlinearity of the MZM form multiple coherent radar subbands at different carrier frequencies. More significantly, these harmonic radar signals facilitate the coherence-processing-free multiband data fusion as a single-channel photonic receiver is able to achieve simultaneous de-chirping processing for multiband radar echoes. In experiments, a 1-GHz-bandwidth transmit LFM signal experimentally obtains a ~8.5-mm range resolution as well as high-resolution inverse synthetic aperture (ISAR) imaging.

## 2. Principle and experimental setup

Figure 1(a) shows the schematic diagram of the proposed photonic multiband radar system. The baseband LFM waveform of narrow bandwidth is generated by a low-speed signal generator and then up-converted to the IF band. This IF LFM signal then drives the MZM through an electrical amplifier (EA). When the driving IF LFM signal is $s(t) = \cos[\phi(t)]$ and $\phi(t) = 2\pi f_\Omega t + \pi k t^2$, due to the modulation nonlinearity of the MZM, its optical output has optical sidebands of different orders:

$$E_t(t) = A_i(m,\theta) e^{j\omega_c t} \sum_{i=0}^{\infty} \cos[i\phi(t)], \qquad (1)$$

where $A_i(m,\theta)$ denotes the amplitudes of the $i$th-order optical sidebands which is related to the modulation index $m$ and bias angle $\theta$ of the MZM [7]. $\omega_c$ is the angular frequency of the optical carrier from a laser diode (LD). $f_\Omega$ and $k$ represent the initial frequency and chirp rate of the input IF LFM signal. After being boosted by an erbium-doped fiber amplifier, the output of MZM is divided into optical paths. One serves as the optical reference signal for photonic radar de-chirping processing. Another one is sent to a photodetector (PD) to generate multiband LFM radar signals. Thanks to the square-law detection in PD, the different orders of harmonics for the input IF LFM signal can be generated. The recovered photocurrent at the output of PD can be written as:

$$I(t) \propto |E_t(t)|^2 = \sum_{q=0}^{\infty} A_q^2(m,\theta)\cos^2[q\phi(t)] + \sum_{r=0}^{\infty}\sum_{s \neq r}^{\infty} A_r(m,\theta)A_s(m,\theta)\cos[r\phi(t)]\cos[s\phi(t)] \approx \sum_{l=0}^{\infty} B_l(m,\theta)\cos[l\phi(t)], \quad (2)$$

in which $q$, $r$, $s$ and $l$ are integers. Accordingly, the recovered multiband RF signal contains the fundamental LFM signal and its harmonics of different orders. Their amplitudes are governed by the bias angle and modulation index of MZM which can be judiciously controlled. Then, the generated multiband LFM radar signals are amplified by an EA and radiated toward the targets through a wideband antenna. For radar ranging demonstrations, two closely spaced 2 cm×3 cm metallic reflectors are used to emulate the targets of interest, see Fig. 1(b). For ISAR imaging verifications, three 2 cm×3 cm metallic targets are mounted on an electric circular rotating platform to emulate dynamically moving targets, see Fig. 1(c).

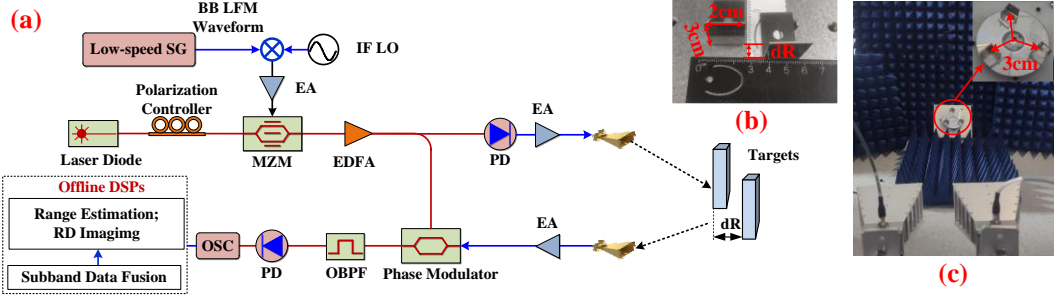

Fig.1. (a) Schematic diagram of the proposed photonic multiband radar system; (b) Photographs of targets for radar ranging and imaging demonstrations. (b) BB: baseband; SG: signal generator; LO: local oscillator; EA: electrical amplifier; EDFA: erbium-doped fiber amplifier; PD: photodetector; OBPF: optical bandpass filter; OSC: oscilloscope; RD: range-Doppler.

Echoes of multiple harmonic subbands reflected by targets are collected by the receiving antenna and sent to a wideband phase modulator (PM) to modulate the optical reference signal. In consideration of the weak echoes, a small-signal modulation assumption is made here. Thus, the optical output of PM can be derived as:

$$E_{PM}(t) \propto E_t(t)e^{j\beta I(t-\tau)} \approx A_i(m,\theta) J_1[B_l(m,\theta)] e^{j\omega_c t} \sum_{i=0}^{\infty}\sum_{l=0}^{\infty} \{e^{j[\pm i\phi(t) \pm l\phi(t-\tau)]}\}, \quad (3)$$

where $\tau$ denotes the round-trip delay of echo with respect to the transmit signal. An optical bandpass filter (OBPF) is used to select upper (or lower) sidebands. Afterwards, the optical signal from the OBPF is sent to a low-speed PD for photonic de-chirping processing. For the particular echo signal of the $l$th subband with an initial frequency $l\Omega$, the low-frequency de-chirped signal from the beating between reference and delayed optical signals around the same sideband can be given as:

$$I_{De}(t) \propto \cos(lk\tau t). \quad (4)$$

Thus, the de-chirped signals corresponding to different radar harmonic subbands are centered at $k\tau, 2k\tau, 3k\tau, ...$, respectively. Therefore, a single-channel photonic receiver achieves simultaneous de-chirping processing for multiband radar echoes here. The obtained temporal waveform at the output of the low-speed PD is digitized by a real-time oscilloscope. Following the methods in [5] and [6], coherent subband data fusion is subsequently implemented to support high-resolution radar ranging and imaging.

## 3. Experiments

Experiments are carried out based on the setup shown in Fig. 1. A 1-GHz-bandwidth input LFM signal ranging from 4.7 GHz to 5.7 GHz is formed by a commercially-available arbitrary waveform generator (AWG) and sent to the MZM. By properly controlling the power of the input LFM signal and the bias angle of MZM, Figs. 2(a) and (b) show the measured spectra of the input LFM signal and generated multiband radar harmonic signals in different frequency ranges of 4.7-5.7 GHz, 9.4-11.4 GHz, 14.1-17.1 GHz, and 18.8-22.8 GHz, resulting in a total bandwidth of 18.1 GHz. For radar-ranging demonstrations, two targets are located ~2 m away from the transmitting antenna. Figures 2(c)-(d) correspondingly give the normalized amplitude spectra of de-chirped signals corresponding to four harmonic radar subbands of different bandwidths when two targets are separated by 15 cm, 5 cm, and 8.5 mm respectively. From Figs. 2(c)-(d), these low-frequency de-chirped signals of four harmonic radar subbands are equally spaced in the frequency domain. As such, they can be easily separated through digital filtering and thereafter support coherent multiband data fusion processing for ultrawideband radar detection [4-6]. Due to the single-channel broadband photonic transceiver generating and processing multiband radar signals, the computationally-intensive mutual-coherence processing for different subbands is averted here. Theoretical range resolutions for the 1-GHz-, 2-GHz-, 3-GHz-, and 4-GHz-bandwidth subbands are 15 cm, 7.5 cm, 5 cm, and 3.5 cm respectively. When two targets are separated by 15 cm, two spectral peaks are observed for the de-chirped signals of all four subbands, see Fig. 2(c). Whilst, when the separation distance between two targets is reduced to 5cm, the de-chirped signals of the 1-GHz- and 2-GHz-bandwidth subbands show only one spectral peak and hence fail to distinguish two targets, see Fig.2(d). A separation distance between two targets as low as 8.5 mm is beyond the detection capabilities of all subbands, see Fig. 2(e). However, by implementing multiband data fusion for four subbands according to the method in [4], millimeter-level ranging resolution can be achieved. Hence, two targets separated by 8.5 mm can be clearly distinguished by multiband data fusion, see Fig. 2(f).

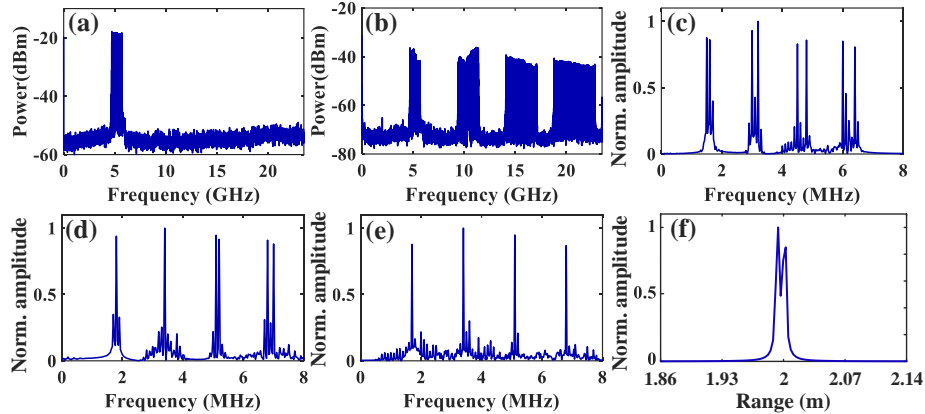

Fig.2. Measured electrical spectra of (a) the input IF LFM signal and (b) generated multiband harmonic radar signals. Normalized amplitude spectra of de-chirped signals for four harmonic radar subbands when two targets are separated by (c) 15 cm, (d) 5 cm, and (e) 8.5 mm. (f) Normalized range profile of four subbands fusion when two targets are separated by 8.5 mm.

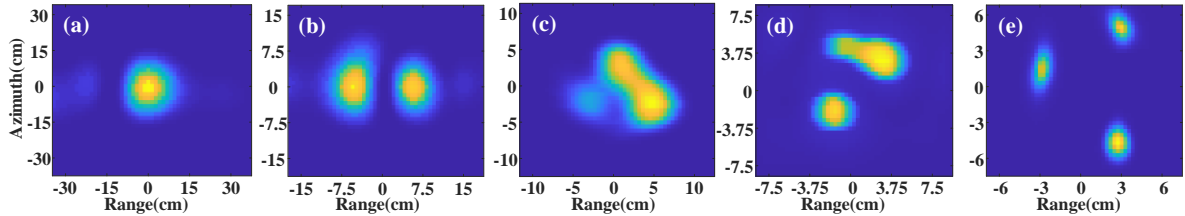

Fig.3. ISAR imaging results based on a single (a) 1-GHz-, (b) 2-GHz-, (c) 3-GHz-, (d) 4-GHz-bandwidth subband and (e) multiband data fusion.

ISAR imaging experiments are further performed based on the experimental setup shown in Fig. 1(c). The rotating platform has a radius of 3 cm and is with a rotation speed of $2\pi$ rad/s. Figures 3(a)-(e) show the imaging results based on a single radar subband of different bandwidths and multiband data fusion. Due to the limited resolution of a single narrow-band sub-band, three targets are not resolvable in both range and azimuth directions, see Fig. 3(a)-(d). However, the data fusion of four subbands enables a noticeably improved ultrawideband radar detection resolution and hence results in a clear observation of three separated targets, see Fig. 3(e).

## 4. Conclusion

A novel ultrawideband photonic multiband radar system is demonstrated here. Using a single MZM and sub-GHz electronics, millimeter-level resolution radar ranging and imaging are experimentally achieved. This work offers a cost-effective and practical solution for meeting ultrahigh-resolution radar detection requirements.